\documentclass[12pt,reprint,amsmath,amssymb,aps,prd,twocolumn,superscriptaddress,floatfix]{revtex4-2}

\usepackage{mathtools}
\usepackage{braket}

\usepackage[caption=false]{subfig}

\usepackage{tikz}
\usepackage[colorlinks, allcolors=blue]{hyperref}
\usepackage{booktabs}
\usepackage{multirow}

\usepackage{csquotes}

\newcommand{\Z}{{\mathbb{Z}}}
\newcommand{\R}{{\mathbb{R}}}

\newcommand{\pqty}[1]{\left( #1 \right)}
\newcommand{\bqty}[1]{\left[ #1 \right]}
\newcommand{\abs}[1]{\left\lvert #1\right\rvert}

\newcommand{\expval}[1]{\langle #1 \rangle}

\newcommand {\dv}[3][ ]{
  \ifx #1 { }
    \frac{d #2}{d #3}
  \else
    \frac{d^{#1} #2}{d #3^{#1}}
  \fi
}

\newcommand {\pdv}[3][ ]{
  \ifx #1 { }
    \frac{\partial #2}{\partial #3}
  \else
    \frac{\partial^{#1} #2}{\partial #3^{#1}}
  \fi
}

\renewcommand{\Re}{\operatorname{Re}}

\newcommand{\Or}{\mathrm{O}}

\newcommand{\U}{\mathrm{U}}

\begin{abstract}

The vortex in the $(2+1)$-dimensional $\Or(2)$ model is studied via numerical simulations in a fully non-perturbative lattice regularization. We compute the vortex condensate and susceptibility to determine its critical exponents and a renormalized condensate in the continuum limit. Together with recent results on the vortex mass, this gives a complete picture of the scaling behaviour of the vortex operator in this model and sheds light on the statistical mechanics of topological excitations.

\end{abstract}

\begin{document}

\title{Vortex condensate and critical exponents in the $(2+1)$-dimensional $\Or(2)$ model}

\newcommand{\affilBern}{\affiliation{Albert Einstein Center for Fundamental Physics, Institute for Theoretical Physics,\\
University of Bern, Sidlerstrasse 5, CH-3012 Bern, Switzerland}}

\author{A. Mariani}
\affilBern

\maketitle

\newpage
 
\section{Introduction}

The $(2+1)$-dimensional $\Or(2)$ model is a quantum field theory which supports vortex excitations. In this context, a vortex is a point-like defect that lives on a two-dimensional spatial timeslice and sweeps a vortex line in the three-dimensional spacetime. This model has a second-order phase transition corresponding to the \enquote{Wilson-Fisher} fixed point, and its Euclidean field theory describes the superfluid transition in $\prescript{4}{}{\mathrm{He}}$ \cite{HeatCapacityLambda, SpecificHeatLambda, Ginzburg1982, XYVicari}. 

Topological excitations may be studied in a Euclidean lattice field theory formulation using the construction of \cite{FrohlichMarchetti87, FrohlichMarchettiVortices}. The vortex in the $(2+1)$-dimensional $\Or(2)$ model can be mapped via an exact duality to the charged particle in an integer-valued lattice gauge theory. This construction leads to a vortex operator which has been rigorously shown to give rise to topological sectors \cite{FrohlichMarchetti87} (note that in the language of \cite{FrohlichMarchetti87} the vortex in this theory would be called a \enquote{monopole}). 

The set-up and methods of this work are the same as those of a previous study \cite{hornung2021mass}, which showed that the vortex constructed via the duality has a mass which is logarithmically divergent with the spatial volume. Near the continuum limit, the $(2+1)$-dimensional $\Or(2)$ Wilson-Fisher fixed point is characterized by a set of critical exponents for the ordinary $\Or(2)$ field, among which those for the mass, $\nu$, magnetization, $\beta$, and susceptibility, $\gamma$. These satisfy a hyperscaling relation,
\begin{equation}
    2\beta + \gamma = \nu d \ , 
\end{equation}
with $d=3$ in this case. The vortex operator will in turn be characterized by a set of critical exponents $\nu_V$, $\beta_V$, and $\gamma_V$ for its mass, condensate (analogous to the magnetization) and susceptibility respectively. Previous work has shown that, within numerical precision, the vortex mass scales with the same exponent as the ordinary mass \cite{hornung2021mass, DelfinoSelke}, i.e. $\nu_V = \nu$. This is consistent with the expectation of a universal scaling for the masses.

In this work, we compute the vortex critical exponents $\beta_V$ and $\gamma_V$ for the condensate and susceptibility. In this case we find that the vortex exponents are significantly different than the usual $\Or(2)$ exponents, i.e. $\beta_V \neq \beta$, and $\gamma_V \neq \gamma$, yet they still satisfy the hyperscaling relation
\begin{equation}
    \label{eq:vortex scaling relation}
    2\beta_V + \gamma_V = \nu_V d \ , 
\end{equation}
where again $d=3$. Such a relation allows the identification of a universal amplitude ratio involving the condensate, susceptibility and mass, which may be alternatively interpreted as a renormalized condensate. 

The critical exponents of the vortex operator are also of interest because in principle they can be directly related to the scaling dimensions of relevant deformations of the conformal field theory at the fixed point. Calculations of scaling dimensions of topological excitations in conformal field theories have attracted attention in recent years \cite{KarthikMonteCarlo, Pufu_2014, Senthil_2019}.

This work, together with \cite{hornung2021mass}, represents the first numerical study of topological excitations using the techniques of \cite{FrohlichMarchetti87, FrohlichMarchettiVortices} in a lattice model with a continuum limit. The same operator has been investigated in the context of monopoles in the $4$-dimensional $\U(1)$ gauge theory \cite{PolleyWiese, JersakNeuhausPfeiffer}, which however only has a first-order phase transition.

The rest of this work is organized as follows. In Section \ref{sec:model} we give a detailed description of the theoretical set-up, in particular the $(2+1)$-dimensional $\Or(2)$ model, the dual theory, and the vortex operator. In Section \ref{sec:numerical}, we discuss the numerical calculation and our main results. 

\section{The model}\label{sec:model}

\subsection{The XY model and its duality}

The $(2+1)$-dimensional $\Or(2)$ model is the theory of a complex scalar field $\Phi(x)$ with Lagrangian
\begin{equation}
    \mathcal{L} = \frac12 \partial_\mu \Phi^* \partial^\mu \Phi - \frac{\lambda}{4!} (\abs{\Phi}^2 - v^2)^2 \ .
\end{equation}
At the classical level, it admits vortex excitations whose energy is logarithmically divergent with the spatial volume. In the quantum theory, for each fixed value of $\lambda$ the system can be driven to the Wilson-Fisher fixed point by tuning $v^2$. In this work we choose to set $\lambda = +\infty$, which has the effect of fixing the absolute value of the scalar field, $\abs{\Phi}^2 \equiv v^2$. The remaining degree of freedom is the angle $\varphi$ defined as $\Phi = v \exp{(i \varphi)}$. One then obtains the Lagrangian of the XY model,
\begin{equation}
    \mathcal{L} = \frac12 v^2 \partial_\mu \varphi \partial^\mu \varphi \ , \qquad \varphi \in [0, 2\pi) \ .
\end{equation}
The theory can be formulated on a three-dimensional Euclidean lattice, in which case the partition function takes the form
\begin{equation}
    \label{eq:original partition function}
    Z = \pqty{\prod_x \frac{1}{2\pi} \int_{-\pi}^{\pi} d\varphi_x} \exp{\pqty{-\sum_{\expval{xy}} s(\varphi_x, \varphi_y) }} \ ,
\end{equation}
where $\expval{xy}$ are nearest neighbours. Several equivalent choices for the action $s$ are possible. Here we choose the Villain form,
\begin{equation}
    \label{eq:villain action}
    e^{-s(\varphi_x, \varphi_y)} = \sum_{n_{\langle x y \rangle} \in \Z} \exp{\bqty{-\frac{1}{2g^2} \pqty{\varphi_x -\varphi_y + 2\pi n_{\langle x y \rangle}}^2}} \ ,
\end{equation}
where $g^2$ is the coupling and the action enforces the $2\pi$ periodicity of the $\varphi$ variables. For small $g^2$, the $\Or(2)$ model is found in a broken phase with a massless Goldstone boson, while for large $g^2$ the $\Or(2)$ symmetry remains unbroken. The two phases are separated by a second-order phase transition which (for this specific choice of action) is accurately known and located at \cite{Neuhaus_2003}
\begin{equation}
    g_c^2 = 3.00239(6) \ .
\end{equation}
In three dimensions, the XY model admits an exact duality transformation to a pure integer-valued non-compact gauge theory. In particular, the dual gauge field $A \in \Z$ lives on the links of the dual lattice, and (up to irrelevant prefactors) the partition function eq.\eqref{eq:original partition function} can be shown to be identical to the partition function
\begin{equation}
    \label{eq:integer gauge theory}
    Z =\prod_{l} \sum_{A_l \in \Z} \exp{\pqty{-\frac{g^2}{2} \sum_{\square} F_\square^2}} \ ,
\end{equation}
where the product is over all links $l$ of the dual lattice, the sum is over all dual lattice plaquettes $\square$ and the non-compact field strength is defined as $F_\square = A_1 + A_2 - A_3 - A_4$ for the four gauge field variables living on the oriented links belonging to plaquette $\square$.

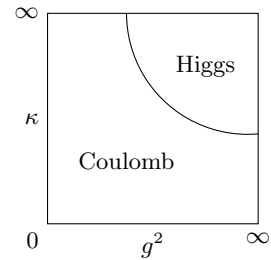
\begin{figure}
    \centering
    \begin{tikzpicture}[scale=0.7]
        \draw (0,0) -- (4,0) -- (4,4) -- (0,4) -- cycle;
        \draw (1.5,4) arc (180:275:2.3cm);

        \node[below left] at (0,0) {$0$};
        \node[below] at (4,0) {$\infty$};
        \node[left] at (0,4) {$\infty$};

        \node[below] at (2,0) {$g^2$};
        \node[left] at (0,2) {$\kappa$};

        \node at (3,3) {Higgs};
        \node at (1.5,1.2) {Coulomb};
    \end{tikzpicture} 
    \caption{Phase diagram of non-compact 3D scalar QED. The integer gauge theory eq.\eqref{eq:integer gauge theory} corresponds to $\kappa = \infty$.}
    \label{fig:scalar qed phase diagram}
\end{figure}

The integer gauge theory eq.\eqref{eq:integer gauge theory} may be interpreted as the $\kappa \to \infty$ limit of a scalar QED theory with a compact scalar $\Phi \in \U(1)$ and a real-valued gauge field $A \in \R$, together with the action
\begin{equation}
    \label{eq:scalar qed}
    S[A,\Phi] = \frac{g^2}{2} \sum_{\square} \pqty{F_\square}^2 - \kappa \sum_{\langle x y \rangle} \Re \pqty{\Phi_x e^{2\pi i A_{\langle x y \rangle}} \Phi_y^*} \ .
\end{equation}
In particular, choosing the unitary gauge $\Phi \equiv 1$ and taking the limit $\kappa \to \infty$, the gauge field becomes integer valued, $A_{\expval{xy}} \in \Z$ and one recovers eq.\eqref{eq:integer gauge theory}. The phase diagram of non-compact scalar QED is shown in Fig.\ref{fig:scalar qed phase diagram}. In particular, for small $g^2$ the integer gauge theory is found in a Coulomb phase (which is therefore dual to the broken phase of the original XY model), while for large $g^2$ the theory is found in a Higgs phase (dual to the symmetric phase). The $\Or(2)$ vortices exist as particles in the Coulomb phase, but they condense in the Higgs phase. 

The duality between the $\Or(2)$ model and the integer gauge theory is not simply an identity between their partition functions, but also provides a dictionary which relates objects in the two theories. For example, the massless photon in the Coloumb phase of the integer gauge theory is dual to the the massless Goldstone boson in the broken phase of the $\Or(2)$ model.

The scalar QED theory eq.\eqref{eq:scalar qed} may be interpreted as the Landau-Ginzburg theory describing the superconducting phase transition \cite{Neuhaus_2003}. In this language, the integer gauge theory eq.\eqref{eq:integer gauge theory} has also been referred to as the \enquote{frozen superconductor} and the Higgs phase is nothing but the superconducting phase.

In this work, we are interested in studying the vortex in the $\Or(2)$ model, which is mapped by the duality to the charged particle in the integer gauge theory. This vortex should not be confused with the \enquote{superconducting} vortex in the Higgs phase of scalar QED, which is a stable particle with a finite mass in the infinite volume limit.

\subsection{The vortex operator}

Under the duality, the vortex in the original XY model is mapped to the charged particle in the integer gauge theory \cite{FrohlichMarchetti87, PolleyWiese, hornung2021mass}. The scalar field in scalar QED is not gauge-invariant. Therefore by itself it is not a valid observable and must be combined with an appropriate factor involving the gauge field (a \enquote{photon cloud}) to make it gauge-invariant. Several choices are possible. Since the vortex is a point-like defect which lives on a spatial timeslice, the correct choice of operator turns out to be equivalent to gauge-fixing to the Coulomb gauge \cite{FrohlichMarchetti87, PolleyWiese, hornung2021mass}. This results in the operator 
\begin{equation}
    \label{eq:full vortex operator}
    \Phi_V(x) = \exp{\pqty{-2\pi i\Delta^{-1} \delta A} } \Phi_x \ ,
\end{equation}
where $\delta$ is the \textit{spatial} lattice divergence and $\Delta$ the \textit{spatial} lattice Laplacian. This is nothing but a lattice version of Dirac's dressed electron field \cite{Dirac}. It is important to emphasize that the operator $\Phi_V$ in eq.\eqref{eq:full vortex operator} is gauge-invariant, and is therefore a valid observable. The exponential in eq.\eqref{eq:full vortex operator} represents the photon cloud which surrounds the charged particle in the spatial directions only, consistently with its interpretation as a vortex. In the limit where scalar QED reduces to the integer gauge theory, the operator \eqref{eq:full vortex operator} reduces to
\begin{equation}
    \label{eq:vortex operator}
    \Phi_V(x) = \exp{\pqty{-2\pi i\Delta^{-1} \delta A} } \ ,
\end{equation}
where now $A$ is integer-valued. In particular, it has been rigorously shown that the operator \eqref{eq:vortex operator} gives rise to topological sectors \cite{FrohlichMarchetti87}, which may be interpreted as vortices. We will therefore refer to $\Phi_V$ as the \enquote{vortex operator}. It should be emphasized that the vortex operator is \textit{non-local} due to its Coulomb cloud. In fact, the vortex is an \textit{infraparticle} \cite{FrohlichMarchetti87, schroer1963infra, Buchholz1982, FrohlichMorchioStrocchi} which is especially infrared sensitive. In the following, we will be interested in computing the vortex condensate, i.e. the condensate of the operator \eqref{eq:vortex operator}. The general problem of whether the scalar field condenses in scalar QED has been considered in the past in mathematically rigorous work in three or higher dimensions, where it was found that the answer is gauge-dependent \cite{KennedyKing1,KennedyKing2,FrohlichHiggs,BorgsNill}. This is however not directly relevant to our case, since we're interested in the specific choice in eq.\eqref{eq:vortex operator}; in this case, condensation has been shown in the relevant phase in four dimensions \cite{FrohlichMarchetti87}, but we are otherwise not aware of results directly applicable to the operator \eqref{eq:vortex operator}.

\begin{table*}[t]
    \begin{center}
    \setlength{\tabcolsep}{4pt}
    \begin{tabular}{ccccccccc}
    \toprule
    \multicolumn{2}{c|}{Both} & \multicolumn{3}{c|}{Coulomb} & \multicolumn{4}{c}{Higgs} \\
    \midrule
    $\abs{g^2-g_c^2}$ & $L_{\mathrm{max}}$ & $g^2$ & sweeps & $\chi_V$ & $g^2$ & sweeps & $\Sigma_V$ & $\xi$ \\
    \midrule
    %0.115499 &  $56$ & 2.886882 & $2 \cdot 10^6$ & 12.8(4) & 3.117898 & $5 \cdot 10^5$ & 0.788(4) & 3.13(2) \\
    %0.064116 &  $56$ & 2.938267 & $2 \cdot 10^6$ & 40(1)   & 3.066513 & $5 \cdot 10^5$ & 0.714(7) & 4.75(2) \\
    0.042051 &  $64$ & 2.960331 & $2 \cdot 10^6$ & 83(2)   & 3.044449 & $5 \cdot 10^5$ & 0.67(1)  & 6.36(2) \\
    0.030270 &  $80$ & 2.972112 & $2 \cdot 10^6$ & 138(4)  & 3.032668 & $5 \cdot 10^5$ & 0.626(6) & 7.97(3) \\
    0.023123 &  $80$ & 2.979259 & $2 \cdot 10^6$ & 223(13) & 3.025521 & $5 \cdot 10^5$ & 0.60(1)  & 9.59(3) \\
    0.018407 & $112$ & 2.983975 & $4 \cdot 10^6$ & 285(9)  & 3.020805 & $1 \cdot 10^6$ & 0.55(1)  & 11.20(4) \\
    0.015103 & $112$ & 2.987278 & $4 \cdot 10^6$ & 419(22) & 3.017502 & $1 \cdot 10^6$ & 0.53(1)  & 12.80(4) \\
    0.012683 & $128$ & 2.989699 & $8 \cdot 10^6$ & 541(24) & 3.015081 & $2 \cdot 10^6$ & 0.51(1)  & 14.41(5) \\
    \bottomrule
    \end{tabular}
    \end{center}
    
    \caption{Parameters and results of the simulations in the Coulomb and Higgs phases. The couplings in the two phases are chosen so that they are at the same distance from the phase transition, which is located at $g_c^2 = 3.00239(6)$. The maximum spatial lattice size is the same for corresponding couplings. The quoted values for $\Sigma_V$, $\chi_V$ and $\xi$ refer to their extrapolation to infinite volume.}
    \label{tab:couplings and co}
\end{table*}

\subsection{Boundary conditions}

A technical problem in the definition of the vortex operator \eqref{eq:vortex operator} is that the Laplacian on a finite lattice with periodic boundary conditions has a zero mode and is therefore not invertible. This is a manifestation of the well-known fact that, since the torus has no boundary, by Gauss' law it cannot support states with non-zero charge. This problem can be overcome by employing $C$-periodic boundary conditions in the spatial directions \cite{PolleyWiese, KronfeldWiese, Wiese1992}. In this case $C$-periodic boundary conditions amount to a charge conjugation transformation when crossing the spatial boundary,
\begin{equation}
    A_{\mu}(x+L \hat{i}) = -A_{\mu}(x) \ .
\end{equation}
In the time direction the boundary conditions remain strictly periodic.

Another technical issue is that, with periodic boundary conditions, due to the non-compactness of the height variables the partition function of the integer gauge theory eq.\eqref{eq:integer gauge theory} is infinite, i.e. $Z=\infty$, even in a finite volume. For a thorough discussion of this problem, see Appendix A in \cite{InfiniteZ}. The source of this infinity is two-fold. First of all, since the partition function is gauge-invariant, the summation over some of the gauge field variables is redundant; due to the non-compact nature of the gauge field, these extra summations then give rise to the infinity. At the formal level, as long as one considers only gauge-invariant observables, as we do, this problem can be resolved by a gauge-fixing scheme which removes the summation over the redundant variables. However, with periodic boundary conditions, this is not sufficient to resolve the problem and the partition function is still infinite. In particular, the partition function eq.\eqref{eq:integer gauge theory} is invariant under the transformation
\begin{equation}
    \label{eq:shift symmetry}
    A_\mu(x) \to A_\mu(x) + n_\mu \ ,
\end{equation}
where $n_\mu$ is an integer which depends only on the direction $\mu$. Importantly, this is not a gauge transformation \cite{InfiniteZ}. Because of this symmetry, even after gauge-fixing the action is independent of one of the gauge field variables (or more, depending on how many directions are periodic), which therefore leads to an unconstrained infinite sum in the partition function eq.\eqref{eq:integer gauge theory}. The transformation eq.\eqref{eq:shift symmetry} is a symmetry of the action with periodic boundary conditions, but not with $C$-periodic boundary conditions; therefore in our case this issue only involves the time direction where we apply periodic boundary conditions. As long as one considers only observables which are invariant under the symmetry eq.\eqref{eq:shift symmetry}, removing the summation over the redundant gauge field variable is then sufficient at the formal level to make the partition function finite. In practice, this issue will not affect the calculation of expectation values of observables which are invariant under eq.\eqref{eq:shift symmetry}, because the summation over the redundant variable will contribute a constant (albeit infinite) prefactor identical in both numerator and denominator, which is therefore irrelevant.

\subsection{Symmetries}

We note that, while the action of scalar QED is generally invariant under non-compact $\U(1)$ gauge transformations, $C$-periodicity explicitly breaks the global $\U(1)$ symmetry down to a $\Z_2$ subgroup which acts as a reflection of the scalar field, $\Phi \to -\Phi$. Under this symmetry, the vortex operator transforms as
\begin{equation}
    \Phi_V \to - \Phi_V \ ,
\end{equation}
and we therefore call this $\Z_2$ symmetry \enquote{vortex field reflection}. Since the \enquote{vortex field reflection} is a remnant $\U(1)$ gauge transformation, the phase in which it is broken should be understood as a Higgs phase with a massive photon. As we will see in the next section, the vortex operator acquires a vacuum expectation value in the Higgs phase, which is therefore characterized by the spontaneous breaking of this $\Z_2$ symmetry, which is a remnant of the $\U(1)$ symmetry. In particular, the second-order phase transition in this model is in the $\Or(2)$ universality class (\textit{not} in the $\Z_2$ universality class); this is also clear by the duality.

\section{Numerical results}\label{sec:numerical} 

\subsection{Set-up and expectations}

The integer gauge theory, eq.\eqref{eq:integer gauge theory}, may be simulated numerically using a standard Metropolis algorithm. The object of interest is the vortex operator correlation function. In particular, we first project the vortex operator to zero momentum,
\begin{equation}
    \Phi_V(t) \equiv \frac{1}{L^2} \sum_{\vec x} \Re \Phi_V(t, \vec x) \ ,
\end{equation}
where $x = (t, \vec x)$. Note that, with $C$-periodic boundary conditions, $C$-even (resp. $C$-odd) operators obey periodic (resp. antiperiodic) boundary conditions. Therefore only the real part of the vortex operator admits a projection to the zero momentum sector (see also \cite{PolleyWiese, hornung2021mass}). We then compute the correlation function
\begin{equation}
    \label{eq:vortex correlation function}
    C(t) \equiv \expval{\Phi_V(0) \Phi_V(t)} \ .
\end{equation}
The quantities of interest can be extracted from the correlation function eq.\eqref{eq:vortex correlation function} by fitting to the spectral decomposition
\begin{equation}
    \label{eq:spectral decomposition}
    \expval{\Phi_V(0) \Phi_V(t)} = \frac{1}{Z} \sum_{n,m} \abs{\expval{n | \Phi_V | m}}^2 e^{-t E_m + (t-T) E_n}  \ .
\end{equation}
In the Coulomb phase we expect a unique vacuum with vortex number zero. In this phase, vortices exist as particles. The Hilbert space also contains other states with zero vortex number, including those corresponding to vortex-antivortex pairs, as well as states with non-zero vortex number. These latter states are those we are interested in, and they can be probed via the spectral decomposition for the vortex correlation function. In finite volume, due to $C$-periodicity only $\Z_2$ vortex field reflection is a symmetry. As a result vortices and antivortices can mix and vortex number, which is an integer-valued quantity in infinite volume, is only conserved modulo $2$.

Therefore, in the Coulomb phase, through the exponential decay of the correlation function of the vortex operator \eqref{eq:vortex operator}, one may extract the energy of the lowest-lying state in the zero-momentum, vortex field reflection-odd sector, which can be naturally identified with the vortex mass $m_V$. This was studied in a previous work \cite{hornung2021mass} and it was found that it behaves as
\begin{equation}
    m_V \sim \abs{t}^{\nu_V} \log{\frac{L}{L_0}} \ ,
\end{equation}
where $L$ is the spatial volume and $t = g^2 - g_c^2$ is the distance from the phase transition. In particular, the vortex mass is logarithmically divergent with the spatial volume (much like in the classical theory) and scales near the continuum limit with a critical exponent $\nu_V$, which is found to be numerically equal to the critical exponent $\nu$ for the mass of the ordinary field, $\nu_V = \nu$ \cite{hornung2021mass, DelfinoSelke}.

\begin{figure*}
    \centering
    \subfloat{\includegraphics[width=0.48\textwidth]{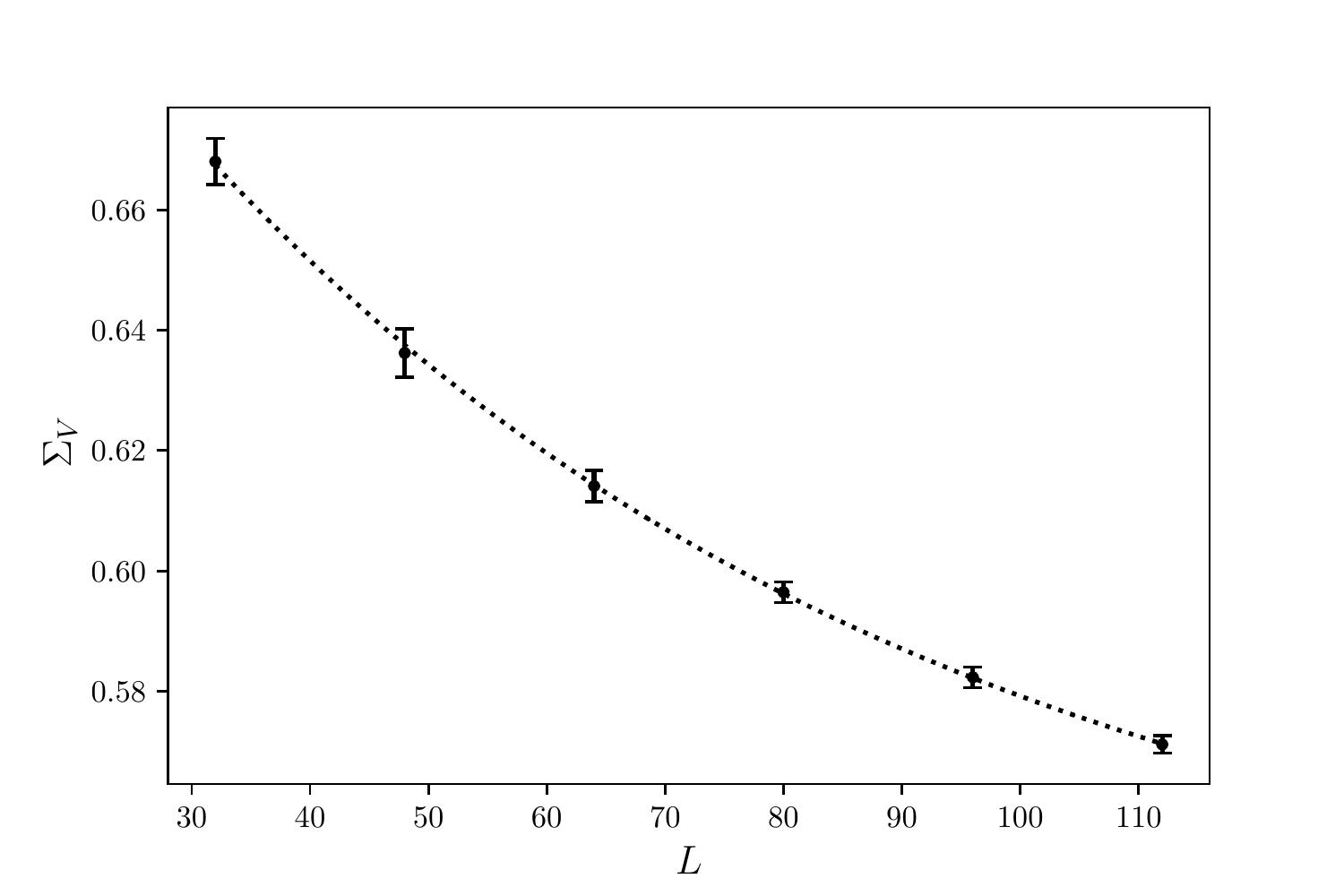}}
    \quad
    \subfloat{\includegraphics[width=0.48\textwidth]{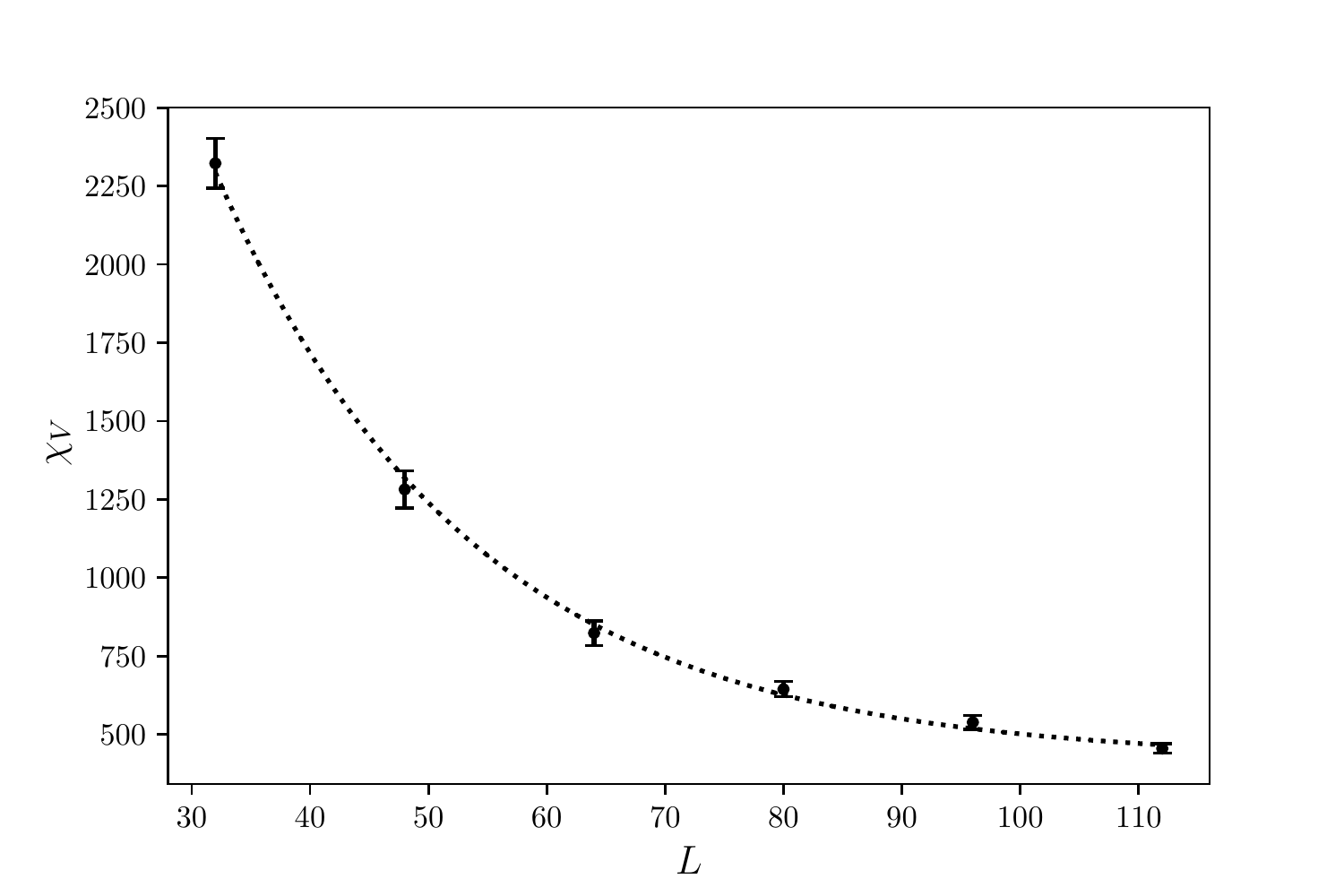}}
    \quad
    \caption{Finite volume results for the condensate (\emph{left}) and susceptibility (\emph{right}) together with an exponential fit at couplings in the two phases at the same distance from the phase transition, with $\abs{g^2-g_c^2}=0.015103$. }
    \label{fig:infinite volume extrapolation}
\end{figure*}

In this work, we consider two other quantities which may be computed from the vortex correlation function: the vortex susceptibility and the vortex condensate. In the Coulomb phase we compute the vortex susceptibility
\begin{equation}
    \label{eq:vortex susceptibility}
    \chi_V \equiv \frac{1}{2 L^2 T} \sum_{x,y} \expval{\Re \Phi_V(x) \Re \Phi_V(y)} \ ,
\end{equation}
where $L^2 T$ is the spacetime volume. This has been defined in analogy with the susceptibility of the ordinary $\Or(2)$ field \cite{Hasenbusch_2008}. In the Coulomb phase the vortices exist as particles and the vortex operator has no vacuum expectation value. Therefore the volume prefactor ensures the correct volume scaling of the susceptibility.

Note that, as can be seen from the spectral decomposition eq.\eqref{eq:spectral decomposition}, the susceptibility eq.\eqref{eq:vortex susceptibility} is well-defined in the infinite volume even though the vortex mass is divergent.

The computation of the susceptibility involves summing the correlation function, eq.\eqref{eq:vortex correlation function}, over $t$ and does not require any fitting. However, since the correlation function decays to zero exponentially while the errors are roughly constant with $t$, in order to avoid artificial accumulation of errors the sum must be truncated. The systematic error arising from the details of the truncation procedure is much smaller than the statistical error of the final result.

Generally speaking, not much is known about the finite volume behaviour or the scaling behaviour of non-local operators such as the vortex operator \eqref{eq:vortex operator}. Therefore, for both the finite volume dependence and the continuum scaling law of the observables we compute, we will assume functional forms which are typical in the scaling theory of local operators, and show that they find numerical confirmation. 

After extrapolating to the infinite volume, we find that the vortex susceptibility scales near the continuum limit with the critical exponent $\gamma_V$,
\begin{equation}
    \chi_V \sim \abs{t}^{-\gamma_V} \ .
\end{equation}

In the Higgs phase, on the other hand, we expect the vortices to condense. This is only possible due to symmetry breaking, which allows the vortex operator to acquire a vacuum expectation value. In finite volume there is no symmetry breaking. Instead, the vacuum is always unique and invariant under the symmetry. Thus in finite volume one finds two almost-degenerate states in the spectrum, $\ket{+}$ and $\ket{-}$, which are respectively even and odd under $\Z_2$ vortex field reflection \cite{JANSEN1988203,JANSEN1989698}. Here $\ket{+}$ is the finite-volume ground state. They form linear combinations 
\begin{equation}
    \ket{0_+} = \frac{\ket{+} + \ket{-}}{\sqrt{2}} \ , \qquad \ket{0_-} = \frac{\ket{+} - \ket{-}}{\sqrt{2}} \ ,
\end{equation}
which are related by vortex field reflection, i.e. $\ket{0_+} \to \ket{0_-}$ and $\ket{0_-} \to \ket{0_+}$. Symmetry breaking implies that these two states become degenerate in the infinite volume limit. The vortex condensate is then given by the absolute value of the vacuum expectation value of the vortex operator,
\begin{equation}
    \bra{0_+} \Phi_V \ket{0_+} = - \bra{0_-} \Phi_V \ket{0_-} \ .
\end{equation}
In finite volume, from the spectral decomposition eq.\eqref{eq:spectral decomposition} we extract the condensate 
\begin{equation}
    \label{eq:vortex condensate definition}
    \Sigma_V \equiv \abs{\expval{+ | \Phi_V | -}} \ ,
\end{equation}
which coincides with the usual definition in the infinite volume limit.  After extrapolating to the infinite volume, we find that the condensate scales near the continuum as
\begin{equation}
    \Sigma_V \sim \abs{t}^{\beta_V} \ .
\end{equation}
As we will see, within numerical precision these critical exponents are found to satisfy the scaling relation $2\beta_V + \gamma_V = \nu_V d$. We can therefore construct the universal amplitude ratio
\begin{equation}
    \label{eq: universal amplitude ratio}
    R_V \equiv \frac{\Sigma_V^2(t) \xi(t)^3}{\chi_V(-t)} \ ,
\end{equation}
which is finite in the continuum limit. Here $\xi$ is the usual (second-moment) correlation length in the symmetric phase of the original $\Or(2)$ model, which scales as $\xi \sim \abs{t}^{-\nu}$. We choose to work with $\xi$ because it is finite in the infinite volume limit, unlike the vortex mass $m_V$. The universal amplitude ratio eq.\eqref{eq: universal amplitude ratio} has been previously considered for the ordinary field in the $\Or(2)$ model \cite{Hasenbusch_2008} and may be equivalently interpreted as a renormalized vortex condensate. An attempt to obtain a renormalized condensate for the same operator had been made in the $4$-dimensional pure $\U(1)$ gauge theory, but it was found not to be possible \cite{JersakNeuhausPfeiffer}. This is likely because the $4$-dimensional pure $\U(1)$ gauge theory only has a first-order phase transition and therefore no continuum limit. 

\begin{figure*}
    \centering
    \subfloat{\includegraphics[width=0.48\textwidth]{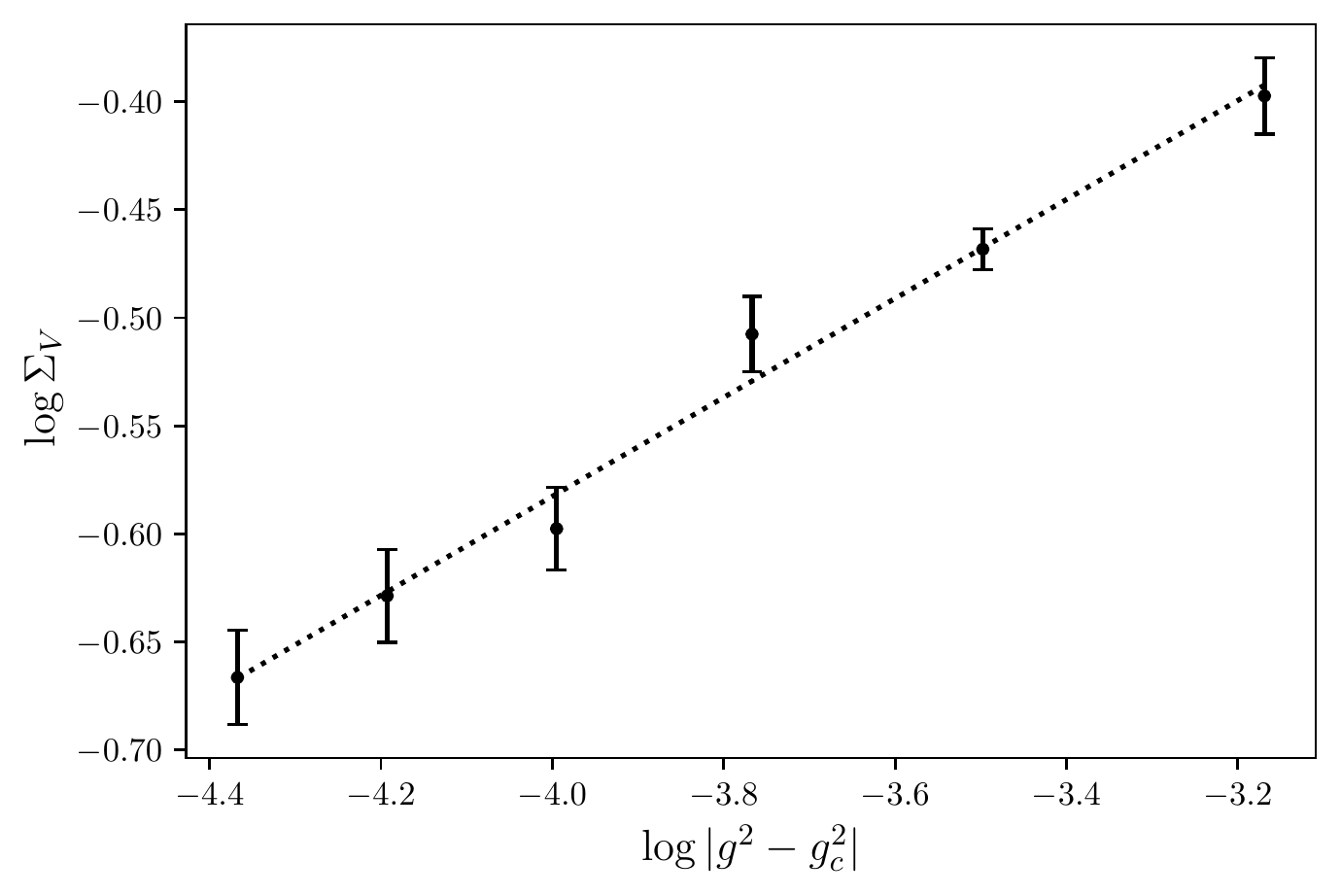}}
    \quad
    \subfloat{\includegraphics[width=0.48\textwidth]{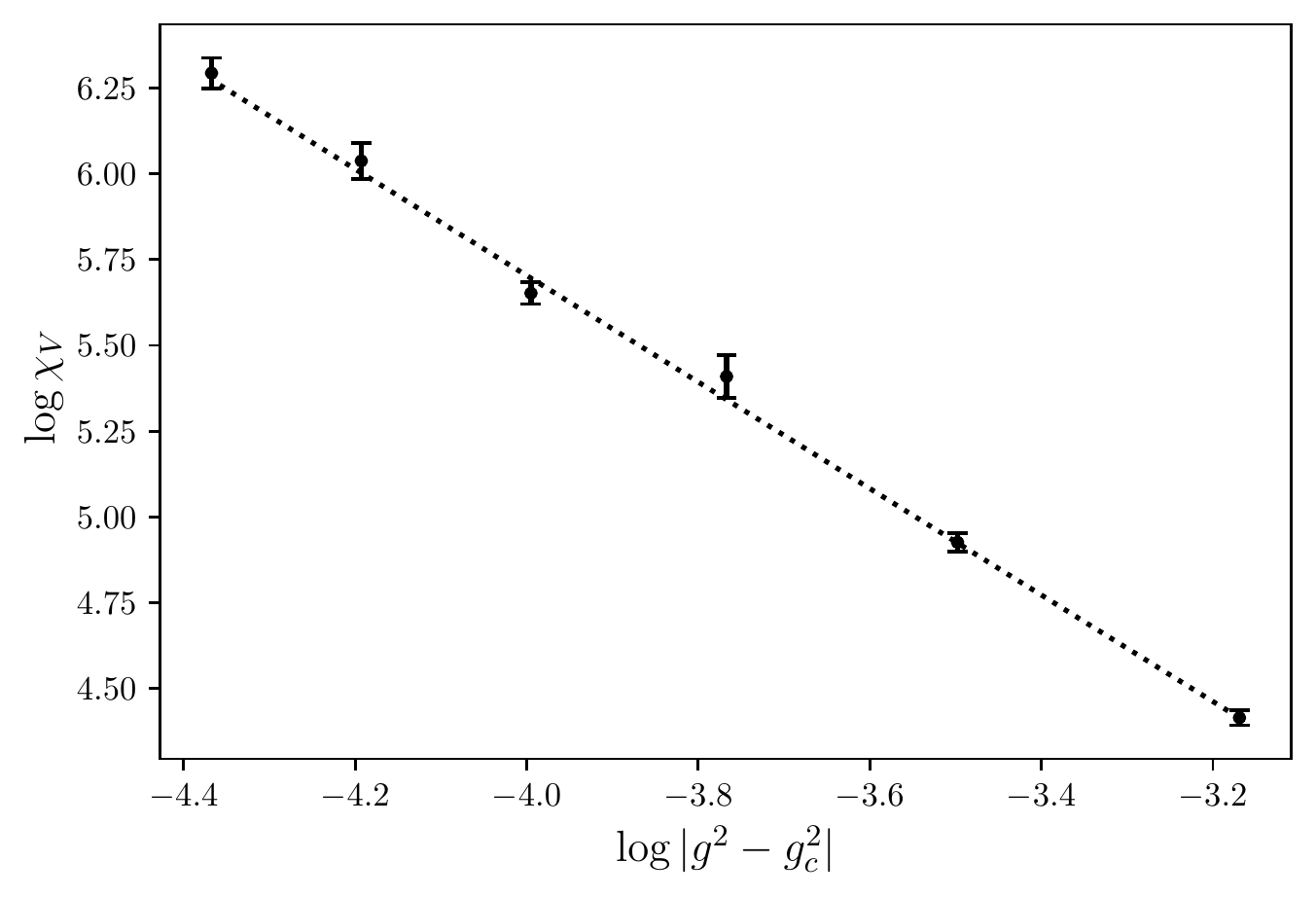}}
    \quad
    \caption{Scaling near the continuum for the condensate (\emph{left}) and  susceptibility (\emph{right}) on a logarithmic scale.}
    \label{fig:continuum extrapolation}
\end{figure*}

\begin{figure}
    \centering
    \includegraphics[width=0.48\textwidth]{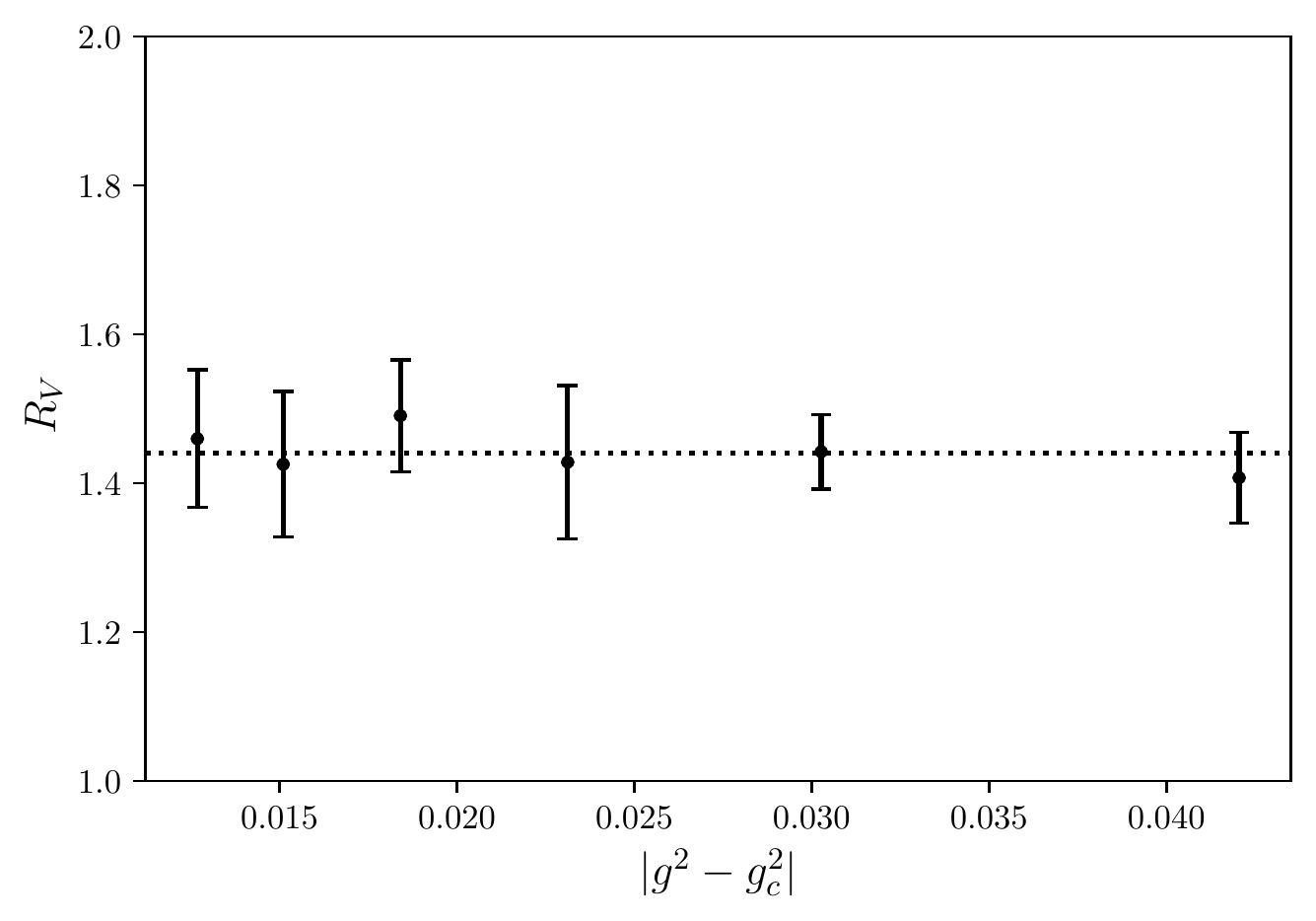}
    \caption{Scaling near the continuum of the amplitude ratio eq.\eqref{eq: universal amplitude ratio}.}
    \label{fig:amplitude ratio}
\end{figure}

\subsection{Results and discussion}

In the Higgs phase, the numerical simulations (for the condensate) were performed at six different couplings $g_i^2$ close to the phase transition, as listed in Table \ref{tab:couplings and co}. Simulations in the Coulomb phase (for the susceptibility) were performed at couplings at the same distance from the phase transition, i.e. at couplings $2 g_c^2 - g_i^2$. For each coupling, the simulations were performed on a lattice of size $L^2 T$ for fixed time extent $T$ and increasing spatial extents $L$. Depending on the coupling, between $6$ and $8$ values of $L$ were simulated. For the susceptibility, simulations with fixed time extent $T=72$ were found to be sufficient, while for the condensate we found a small but significant error from the extrapolation to infinite $T$. We have therefore repeated each simulation for $5$ increasing values of $T$ and included the difference in the error budget. A summary of the couplings, largest $L$, and number of Metropolis sweeps is presented in Table \ref{tab:couplings and co}. Measurements were taken every $100-400$ sweeps depending on the vicinity to the phase transition, and then further appropriately binned. Note that more statistics are required in the Coulomb phase in order to distinguish the exponential behaviour of the finite volume susceptibility (as opposed to other plausible fitting forms such as power laws). 

Both the susceptibility and condensate were then extrapolated to the infinite volume by fitting to the exponential form
\begin{equation}
    \label{eq:infinite volume correction}
    f(L) = f(\infty) + A e^{-B L} \ ,
\end{equation}
where $f(\infty), A$ and $B$ are free parameters. Perhaps surprisingly,  we find exponentially small finite-volume corrections in both the Higgs phase and Coulomb phases. For the condensate, the infinite volume fits have $\chi^2/\mathrm{d.o.f.}$ in the range $0.26-1.04$ with the exception of the second-closest coupling to the phase transition, for which $\chi^2/\mathrm{d.o.f.} \approx 0.06$, while for the susceptibility the $\chi^2/\mathrm{d.o.f.}$ ranges between $0.65-1.11$. Example fits to the finite volume data for both quantities are shown in Fig.\ref{fig:infinite volume extrapolation}. The susceptibility overall has much stronger finite volume effects. The quality of our results could be improved by a better theoretical understanding of the finite volume effects. The final results of the extrapolation to infinite volume are also given for each coupling in Table \ref{tab:couplings and co}.

Finally, we extrapolate to the continuum by fitting to a straight line on a logarithmic scale, as shown in Fig.\ref{fig:continuum extrapolation}. The fit for the condensate has a $\chi^2/\mathrm{d.o.f.}\approx 0.59$, while the one for the susceptibility has $\chi^2/\mathrm{d.o.f.}\approx 0.93$. This allows us to extract the amplitude and critical exponents for both quantities. In Fig.\ref{fig:amplitude ratio} we also plot the amplitude ratio, eq.\eqref{eq: universal amplitude ratio}. The amplitude ratio requires knowledge of the correlation length $\xi$ at the couplings in Table \ref{tab:couplings and co}. This we obtained by fitting the expected scaling form together with corrections to scaling \cite{Hasenbusch_2008} to raw data computed using the Wolff cluster algorithm in the original $\Or(2)$ model and kindly provided by the authors of \cite{hornung2021mass}. We see that the amplitude ratio is consistent with a constant, as expected if the vortex critical exponents indeed satisfy eq.\eqref{eq:vortex scaling relation}. Fitting to a straight line also gives a result consistent with a constant within error.

The final results are therefore found to be
\begin{align}
    \chi_V &= C_\chi \abs{t}^{-\gamma_V} \ , &C_\chi = 0.61(7)  \ ,\quad &\gamma_V = 1.55(3) \ , \nonumber \\
    \Sigma_V &= C_\Sigma \abs{t}^{\beta_V} \ , &C_\Sigma = 1.40(9)  \ , \quad &\beta_V = 0.23(2) \ , \nonumber \\
    \quad & \quad &R_V= 1.44(3) \ . \quad & \quad
\end{align}
The critical exponents can be compared with the values for the ordinary field of the $\Or(2)$ model \cite{Hasenbusch_2019} 
\begin{equation}
    \nu = 0.67169(7)  \ ,  \quad \gamma = 1.378(1) \ , \quad \beta = 0.34864(5) \ ,
\end{equation}
where $\beta$ and $\gamma$ have been obtained from the published values of $\eta$ and $\nu$ in \cite{Hasenbusch_2019} via scaling relations. While we report the ordinary critical exponents for reference, we do not expect the vortex exponents to the be equal to the ordinary ones. In fact, there's no a priori reason why critical exponents associated to different operators should be the same, even in the same universality class \cite{HasenbuschVicari}.

Overall we find that, within numerical precision, the vortex critical exponents satisfy the scaling relation \eqref{eq:vortex scaling relation}:
\begin{align}
    %2 \beta_V + \gamma_V &= 2.009(31) \ , \\
    2 \beta_V + \gamma_V &= 2.01(3) \ , \\
    3\nu & = 2.0151(2) \ .
\end{align}
Since the scaling relation is satisfied, the amplitude ratio \eqref{eq: universal amplitude ratio} is constant in the continuum limit. This can also be seen from Fig. \ref{fig:amplitude ratio}.

The precision of our results is limited by a number of factors. The vortex operator is expensive to compute numerically and since it is non-local, it has less favourable finite-volume scaling than the ordinary $\Or(2)$ field. Moreover, the lack of theoretical understanding of the finite-volume behaviour and the corrections to continuum scaling for the vortex operator limit the precision of the various fits. It is unlikely that significantly better precision can be achieved simply by increasing the statistics.

The vortex operator eq.\eqref{eq:vortex operator} is a non-local infraparticle, and it is therefore unclear whether standard RG theory applies to it. If it can be interpreted as an RG operator, then the standard scaling hypothesis (see for example \cite{PelissettoVicari, HasenbuschVicari}) predicts that the scaling relation eq.\eqref{eq:vortex scaling relation} holds for the vortex critical exponents, and $\nu_V = \nu$. This is indeed what we find in our numerical study, together with the results of \cite{hornung2021mass}. From the same considerations, we can also extract the scaling dimension of the vortex operator, which we find to be 
\begin{equation}
    \Delta_V = \frac{\beta}{\nu} = 0.34(3) \ .
\end{equation}
Since this operator is non-local and the vortex has infinite mass in the infinite volume limit, it is unclear whether it admits a meaningful interpretation in the CFT describing the Wilson-Fisher fixed point. This provides motivation for further investigations in the context of conformal field theories.

\section{Conclusions}

In this work we have computed the critical exponents for the condensate and susceptibility of the vortex in the $(2+1)$-dimensional $\Or(2)$ model. Together with \cite{hornung2021mass}, this is the first time that a non-local infraparticle is investigated in a model which admits a continuum limit. Our work shows that the vortex critical exponents obey a scaling relation, which allows the identification of a renormalized vortex condensate in the continuum limit. 

The results of this work may in principle be verified experimentally by studying vortex lines in three-dimensional superfluids near the Wilson-Fisher fixed point. Moreover, our results show that the construction of \cite{FrohlichMarchetti87, FrohlichMarchettiVortices} may be employed to study topological excitations in the same way as ordinary operators in statistical mechanics models \cite{HasenbuschVicari}. It would also be interesting to better understand the vortex operator in the context of conformal field theories, especially in light of our result on the scaling dimension.

Finally, it would be worthwhile to also extend the results of this work to other models which support vortex excitations, such as the $\U(1)$ Abelian Higgs model or the quantum XY model.

\section*{Acknowledgments}

The author would like to thank U.-J. Wiese, J. C. Pinto Barros, and M. Hornung for useful discussions, and for providing their data on the correlation length. The research leading to this work has received funding from the Schweizerischer Nationalfonds (grant agreement number 200020\_200424).

\bibliography{biblio}

\end{document}